\documentclass[12pt,preprint]{aastex}

\slugcomment{to appear in ApJ, accepted April 18 2011}

\shorttitle{New White Dwarfs in UKIDSS}
\shortauthors{Leggett et al.}

\begin{document}


\title{Cool White Dwarfs Found \\ in the UKIRT Infrared Deep Sky Survey}


\author{S. K. Leggett\altaffilmark{1}}
\author{N. Lodieu\altaffilmark{2,3}}
\author{P.-E. Tremblay\altaffilmark{4}}
\author{P. Bergeron\altaffilmark{4}}
\and
\author{A. Nitta\altaffilmark{1}}

\altaffiltext{1}{Gemini Observatory, Northern Operations Center, 670
  N. A'ohoku Place, Hilo, HI 96720, USA}
\altaffiltext{2}{Instituto de Astrof\'isica de Canarias (IAC), C/ V\'ia L\'actea s/n,
E-38200 La Laguna, Tenerife, Spain}
\altaffiltext{3}{Departamento de Astrof\'isica, Universidad de La Laguna (ULL),
E-38205 La Laguna, Tenerife, Spain}
\altaffiltext{4}{D\'epartement de Physique, Universit\'e de Montr\'eal,
C.P.\ 6128, Succursale Centre-Ville, Montr\'eal, QC H3C 3J7, Canada}


\begin{abstract}
We present the results of a search for
cool white dwarfs in the United Kingdom InfraRed Telescope (UKIRT)
Infrared Deep Sky Survey (UKIDSS) Large Area Survey (LAS). The UKIDSS
LAS photometry was paired with the Sloan Digital Sky Survey (SDSS) to identify
cool hydrogen-rich white dwarf candidates by their neutral optical colors
and blue near-infrared colors, as well as faint Reduced Proper Motion magnitudes. Optical spectroscopy was obtained at
Gemini Observatory, and showed the majority of the candidates to be newly identified cool degenerates, with a small number of G- to K-type (sub)dwarf contaminants. Our initial search of 280 deg$^2$ of sky resulted in seven new
white dwarfs with effective temperature  $T_{\rm eff} \approx$ 6000~K. The current followup of 1400 deg$^2$ of sky has produced thirteen new white dwarfs. Model fits to the photometry show
that seven of the newly identified white dwarfs have 4120~K $\leq T_{\rm eff} \leq$ 4480~K, and cooling ages between 7.3 Gyr and 8.7 Gyr; they have 40 kms$^{-1} \leq v_{\rm tan} \leq$ 85 kms$^{-1}$ and are likely to be thick disk 10--11 Gyr-old objects.
The other half of the sample has 4610~K $\leq T_{\rm eff} \leq$ 5260~K, cooling ages between 4.3 Gyr and 6.9 Gyr,
and 60 kms$^{-1} \leq v_{\rm tan} \leq$ 100 kms$^{-1}$. 
These are either thin disk remnants with
unusually high velocities, or lower-mass remnants of 
thick disk or halo late-F or G stars.

\end{abstract}




\section{Introduction}

White dwarfs are the end stage of stellar evolution for the vast majority of
stars --- all stars with initial mass between around $0.07 M_{\odot}$ and $8  M_{\odot}$
end their lives as  cooling white dwarfs.
The cooling rate of these degenerate remnants slows as their temperature drops,
such that very old white dwarfs are still visible.
The coolest white dwarfs can constrain the age of the Galactic disk, or even of the
halo (e.g. Winget et al. 1987).  
While other methods for dating the Galactic components
exist ---
asteroseismology, gyrochronology, isochrone fitting, isotope decay,  magnetic activity, kinematics and metallicity (e.g. Ulrich 1986, Barnes 2007, Chaboyer et al. 1996, Frebel et al. 2007, West et al. 2008, Nordstrom et al. 2004, see also the review Soderblom 2010) ---
very cool white dwarfs are unambiguously old and their cooling rates are quite well understood, enabling
their use as accurate chronometers (e.g. Iben \& Tutukov 1984; Fontaine, Brassard \& Bergeron 2001).
Most white dwarfs consist of a C/O core with an outer
envelope composed of helium and/or hydrogen, with occasional traces of metals. The
mass and composition of both the core and the atmosphere are important in determining the age of a 
white dwarf. The mass and core composition determine the thermal content, while the composition of the atmosphere provides the insulating material which
controls the rate of cooling.  Bergeron et al. (2001) use atmospheric and evolutionary models to analyse a sample of white dwarfs with measured trigonometric parallaxes to show that the coolest of these 
white dwarfs, with $T_{\rm eff} \sim$4000--4500~K, are 9--10 Gyr old 
if they have a thick hydrogen atmosphere, and 8--9 Gyr old if they have a 
helium-rich atmosphere. These ages are consistent with the age of the 
local Galactic disk (e.g. Leggett et al. 1998).

Several groups are trying to find old white dwarfs to confirm the age of the disk and to investigate the
ages of the older Galaxy components. Candidate white dwarfs are typically identified as high proper motion
objects, or by their colors or spectra, or by a combination of their kinematic and photometric properties.
Using kinematic data alone is problematical. Oppenheimer et al.\ (2001) identified a sample of 
high-velocity white dwarfs which was inferred to be a halo population by 
their kinematics. However Reid et al. \ (2001) suggest that the majority of this sample has kinematics consistent with thick disk membership, and analysis of the sample by Bergeron et al.\ (2005) found that the white dwarfs were relatively warm, implying relatively short cooling ages. The short cooling age does not necessarily exclude the possibility that these white dwarfs are old --- they may have evolved from low-mass stars with long main-sequence lifetimes.

Omitting infrared data in studies of candidate cool white dwarfs can also lead to large uncertainties in the derived $T_{\rm eff}$ and hence age.  This is because the complete optical to infrared spectral energy distribution is required to determine the atmospheric composition and $T_{\rm eff}$. Hydrogen-rich white dwarfs cooler than about 5000~K (depending on  instrument resolution) are featureless, however the infrared region demonstrates the presence of hydrogen due to the opacity of pressure-induced H$_2$ absorption
(e.g. Borysow 2002).
Kilic et al. (2010a) perform a detailed analysis of cool white dwarfs using optical spectroscopy and
infrared photometry. They determine significantly different values of $T_{\rm eff}$ from those found using
optical data only; several white dwarfs thought to have $T_{\rm eff} \sim$3500 K in fact have $T_{\rm eff} \sim$4500 K, and the cooling age
decreased from $>$10~Gyr to $<$8~Gyr.
    
Very cool white dwarfs are unambiguously old,
as their total age is dominated by the large, post-main sequence, cooling time.
There are about 20 white dwarfs known with $T_{\rm eff} < 4000$~K. These have been found as high proper-motion objects in
photographic sky surveys (Hambly et al. 1997, Harris et al. 1999, Ibata et al. 2000, Oppenheimer et al. 2001, Rowell et al. 2008, Ruiz \& Bergeron 2001, Scholz et al. 2002) or in the Sloan Digital Sky Survey
(SDSS, York et al. 2000) by inspection of spectra or by using the Reduced Proper Motion (RPM; Gates et al. 2004 , Hall et al. 2008, Harris et al. 2001, Harris et al. 2008, Kilic et al. 2006, Kilic et al. 2010b).
The RPM is defined as: 
$$ H = mag + (5\times log(PM) + 5) $$
where $PM$ is the proper motion in $\arcsec$ yr$^{-1}$, and the RPM acts as a proxy for absolute magnitude for a sample with similar kinematics.
Faint white dwarfs with significant proper motion can be separated from stars and subdwarfs using an RPM diagram, such as that shown in Figure 1.
       
Our group has paired the UKIRT Infrared Deep Sky Survey (UKIDSS, Lawrence et al. 2007) with the SDSS, 
to provide a full optical through near-infrared spectral energy distribution,
and to select candidate cool white dwarfs using the RPM and various colors. 
The proper motions are derived using the target coordinates and the epochs of the SDSS and UKIDSS images.
Combining these databases allows us to go about a magnitude deeper in the $g$-band than
other searches, to $g \approx 21$, although we are also limited to $r \approx 21$ by the time required to obtain a spectrum
for classification.
In Lodieu et al. (2009b) we published the results of a search of 280 deg$^2$ of the Large Area Survey (LAS) 
component of the UKIDSS Data Release (DR) 2. The search produced seven new white dwarfs with $T_{\rm eff} \approx 6000$~K, 
which were confirmed with optical spectroscopy obtained at Gemini Observatory.
Here we report the identification and confirmation of white dwarfs selected from 1400  deg$^2$ of DR 6 of the LAS.
The larger area and better understanding of the photometry has allowed us to discover significantly cooler and older
white dwarfs compared to the Lodieu et al. sample.

\section{Sample Selection}

\subsection{The LAS and SDSS Databases}

UKIDSS consists of five survey components, one of which is the LAS (Lawrence et al. 2007). The LAS is the sub-survey most likely to contain 
faint and rare sources of the local Galaxy, such as the cool white dwarfs 
and brown dwarfs.  The LAS aims to survey 4000 deg$^2$ in $YJHK$ (Hewett et al. 2006)
with a second epoch at $J$, to reach $J \sim$ 20. 
The $5 \sigma$ photometric depths are $Y=20.2$, $J=19.6$, $H=18.8$ and $K=18.2$ (Dye et al. 2006).
The data and catalogs generated by the automatic pipeline processing can
be retrieved through the WFCAM Science Archive (WSA; Hambly et al.\ 2008).
All data are pipeline-processed by the Cambridge Astronomical Survey Unit
(CASU) following a standard procedure for
infrared images (Irwin et al. 2004). An extensive description of each step involved in the 
processing of the WFCAM data is available on the CASU 
webpage\footnote{http://casu.ast.cam.ac.uk/surveys-projects/wfcam/technical}.

The work presented here involves spectroscopic follow-up during Gemini Observatory semesters 2008B and 2010A (see \S 3).
Our selection of candidates evolved over time, as LAS Data Releases became available, and as we obtained
spectra for the candidates. The Observatory was flexible in allowing us to replace targets listed
in the proposals with better candidates as they were found in new Data Releases. 
The LAS Releases used were DR3 (December 2007, 809 deg$^2$), DR4 (July 2008, 984 deg$^2$),
DR5 (April 2009, 1270 deg$^2$) and DR6 (October 2009, 1434 deg$^2$).

We have cross-correlated the LAS with the SDSS database,
using Structured Query Language (SQL) and the  WSA.
SDSS Data Releases 6 and 7 were used, as available (Adelman-McCarthy et al. 2008, Abazajian et al. 2009).
The area surveyed 
by the LAS was designed to overlap with the SDSS, divided up into three
blocks (Dye et al. 2006, Lawrence et al. 2007,  Warren
et al. 2007). The equatorial block with  Right Ascension 22 to 04 hours 
and Declination between $-$1.5 and $+$1.5 degrees overlaps SDSS 
stripes 9 to 16. The southern block covers 8 to 16 hours 
and  (approximately) $-$3 to $+$15 degrees, and includes  
SDSS stripe 82. The northern block  covers  8 to 17 hours 
and  (approximately) $+$30 to $+$50 degrees, and provides an overlap 
with SDSS stripes 26 to 33. 

\subsection{Astrometry and Photometry}

Sources were matched by requiring the presence of a
``primary'' SDSS source within 5$\arcsec$ of the LAS coordinates (increasing the search radius led to erroneous pairings). 
The WFCAM astrometry is  tied to the
2MASS point source catalog (Skrutskie et al. 2006)
and has a systematic accuracy of $< 0\farcs 1 
\ rms$ (Dye et al. 2006). Lodieu et al. (2009) plot the $rms$ of the difference between the coordinates of all point sources in the LAS DR2 and SDSS DR5 as a function of $J$ magnitude; for the sources in this sample with $17.5 \leq J \leq 19.2$ the implied astrometric uncertainty is 0$\farcs$025 to 0$\farcs$035 for
the brighter to fainter sources, respectively.

We have restricted our queries of the LAS database to detections classified
as point sources, and imposed color criteria and lower limits on the $g$-band RPM, $H_g$.
The queries return coordinates, 
photometry and errors from both surveys, as well as the proper motion, computed
from the difference between the LAS $J$-band coordinates and the SDSS $z$-band coordinates.
For the sample presented here, the SDSS epoch ranges from 1999 to 2006, and the LAS from 2005 to 2008.
The average time period between the LAS and SDSS astrometry is $4.5\pm2.0$ years. Given the astrometric
uncertainty, the uncertainty in proper motion is $\lesssim$ 14 mas.  For one source, 
ULAS J1323$+$12, the LAS and SDSS epochs are separated by only 0.89 yr and the proper motion uncertainty is 30 mas.

Table 1 gives astrometric information for the candidate white dwarfs studied here, and Table 2 gives the
SDSS $ugriz$ and the LAS $YJHK$ photometry.  Note that the photometry is taken from the 8th Data Releases for both the LAS and SDSS, and not the releases used for our initial source identification. The proper motion in Table 1 is similarly an updated value, and has been calculated from the coordinates and epochs given in the DR8 releases.

With the exception of the close binary ULAS J1234$+$06,
Table 1 also lists proper motions  derived from alternative catalog matching. For 10 of the 16 sources, proper motion derived by 
matching the USNO-B catalog (Monet et al. 2003) to the SDSS catalog is available via DR8 of the SDSS. Of these ten, there
is good agreement for six objects. For two the differences are at the $\sim 2\sigma$ level (ULAS J0121$-$00 and ULAS J1142$+$00).
For the remaining two objects the discrepancies are large, and we rederived the proper motions using the digitised sky images, as well as the USNO-B, SDSS and LAS astrometry, in order to obtain the motion over a large timeline. For  
ULAS J1323$+$12, which has a small SDSS-LAS epoch difference, the derived motion is close to the USNO-SDSS value, and
significantly smaller than the SDSS-LAS value. For ULAS J2246$-$00, the derived value is close to the SDSS-LAS value but significantly different from the USNO-SDSS value in direction. For the six objects without USNO-SDSS proper motions, three show good agreement between the SDSS-LAS proper motions and new measurements derived from the imaging and catalog data listed above. Three objects differ by  $\sim 2\sigma$ (ULAS J0840$+$05, ULAS J1345$+$15, ULAS J1454$-$01). Table 1 gives the RPM values derived from both proper motion measurement sets.  

\subsection{Color and RPM Selection}

Figures 1 and 2 illustrate the selection criteria imposed to arrive at the candidate list of Table 1.
For all queries, we selected $14 \leq J \leq 19.6$, to avoid saturated sources, and to ensure a good ($5 \sigma$)
$J$-band detection, in the LAS. We also restricted the targets to $r_{AB} < 20.7$, so that the spectroscopic
observations were of reasonable duration (see \S 3.1).  To select for faint high-proper motion sources, we
restricted sources to $H_g > 20.5$ (see Figure 1). This last selection greatly reduced the number of objects 
returned from a query. Figure 3 shows the red to far-red colors of various samples, for completeness.

Most sources, 9 of the 17 listed in Table 1, were found in the search identified as region A in Figures 1 and 2, and in Table 1.
The cuts imposed, in addition to those listed above, were: 
$ H < 18.9$, $0.8 \leq g-r \leq 1.6$, $0.2 \leq r-i \leq 0.6$, $0.6 \leq i-J \leq 1.4$, $J-H \leq 0.2$.
This selection is designed to produce hydrogen-rich white dwarfs with $T_{\rm eff} \sim 4000$~K (Figure 2), with a good detection in the $H$-band.
Early on in this work an additional source on the blue side of this cut with $g-r = 0.63$ was followed up;
later we prioritized redder (in $g-r$) sources in order to find white dwarfs cooler than 5000~K. This bluer source is identified
as found in region Ab in Table 1.
An additional 6 sources were found in search B, defined as: $0.2 \leq g-r \leq 1.2$, $-0.6 \leq r-i \leq +0.6$, $J-H \leq -0.1$, $H-K \leq -0.1$. This search is designed to find bluer sources where H$_2$ opacity is impacting the red as well
as the near-infrared, due to either lower $T_{\rm eff}$, or a higher-pressure mixed H-He atmosphere (Figure 2).
Finally, one more source was found by selecting for extreme sources in the RPM diagram (Figure 1): $H_g > 22.5$,  
$-0.5 \leq g-i \leq 1.0$, $0.0 \leq r-i \leq 0.6$, $H < 18.8$, $J-H \leq 0.1$, $H-K \leq 0$. This source is identified
as found in region C in Table 1 (later the RPM value was found to be erroneous).

Our spectroscopic followup is complete for the pairing of the 1270 deg$^2$ of the LAS DR5 and the SDSS DR7 with the following selection:
$ H_g > 20.5$, $r < 20.7$, $14 \leq J \leq 19.6$, $H < 18.9$, $0.9 \leq g-r \leq 1.6$, $0.2 \leq r-i \leq 0.6$, $0.6 \leq i-J \leq 1.4$, 
$J-H +(\sqrt{err(J)^2 + err(H)^2}) < 0.2$. 
This search of the 1270 deg$^2$ of LAS DR5 produced six new white dwarfs with 4120~K $\leq T_{\rm eff} \leq$ 4380~K (see \S 4), and recovered a known white dwarf with $T_{\rm eff} = 4390$~K (SDSS J115814.52+000458.3, Kilic et al. 2010a).  Kilic et al. (2010a) perform a detailed analysis of 113 white  
dwarfs with $v_{\rm tan} \geq 20$~kms$^{-1}$ found using an RPM search of the 5282 deg$^2$ footprint of the SDSS DR3 (Harris et al. 2006). We can compare our results to those of Kilic et al., 
neglecting the additional 13 white dwarfs in the Kilic et al. sample that were found through a search of SDSS spectral data, in order to compare similarly determined samples.
Kilic et al. find 6 white dwarfs with
4100~K $\leq T_{\rm eff} \leq$ 4400~K out to a distance of about 55~pc. The implied space density of 4200~K white dwarfs, with cooling age $\approx 8$~Gyr (\S 4), is $2\times10^{-5}$ pc$^{-3}$. The $r$-, $J$- and $H$-band cuts imposed for our LAS search indicate that we should be sensitive to 4200~K white dwarfs at distances of 10 -- 95~pc, based on the models. The $H_g$ selection combined with a faint limit of $g \approx 21$ implies that our lower limit on proper-motion is 0$\farcs$08 yr$^{-1}$. Our upper limit is determined by the 5$\arcsec$ matching radius and the difference between the SDSS and LAS epochs; the typical difference is four years implying an upper limit to the proper motion of $\sim 1\farcs$3 yr$^{-1}$. Excluding 
ULAS J1142$+$00 and ULAS J1234$+$06AB for which erroneously high proper motions were initially derived,
the sample  has a range in proper motion of 0$\farcs$09 to 0$\farcs$3 yr$^{-1}$.
The density of 4200~K white dwarfs implied by the Kilic et al. sample suggests that we would expect to find around seven 4200~K white dwarfs. However two of the six Kilic et al. white dwarfs have $ H_g < 20.5$, suggesting that our selection would find around five 4200~K white dwarfs, very close to the number found. 

We note in passing that we have applied this last selection to the 48 deg$^2$ overlap area of  the UKIDSS Galactic Cluster Survey (GCS) DR8 and SDSS DR7, and found no candidates. The GCS will survey ten open star clusters in the $ZYJHK$ filters, to a similar depth in $YJHK$ as the LAS. It has been successfully used to find field brown dwarfs (Lodieu et al. 2009a), as well as to study open clusters (Lodieu et al. 2007), but a larger area in common with the SDSS is required before cool white dwarfs will be found.

\section{Observations}




\subsection{Optical Spectroscopy}

We obtained optical spectra of the candidate cool white dwarfs listed in Table 1,
selected as described in the previous Section, excluding  
SDSS J1247$+$06 for which Kilic et al. 2010a present an optical spectrum (a spectrum is also available in
the SDSS database). The Gemini Multi-Object Spectrographs
(GMOS, Hook et al. 2004) at both Gemini North and South were used, through queue time granted under programs
GS-2008B-Q-35, GN-2008B-Q-111 and GS-2010A-Q-58. For all observations the
R400 grating was used with the GG455 blocking filter.  The central wavelength was 680~nm, with wavelength coverage of 
460 -- 890 nm. Detector fringing affected the spectra longwards of about 800 nm. We did not correct for the fringing as this wavelength region is not necessary for target classification.
The 0$\farcs$75 slit was used with $2\times 2$ binning, and the resulting resolution was $R \approx$ 1280 or 6 \AA.
An observing log is given in Table 3.

Flatfielding and wavelength 
calibration were achieved using lamps in the on-telescope calibration unit. 
Spectrophotometric DA white dwarf standards were used to determine the instrument response curve, 
and flux calibrate the spectra; for program  GS-2008B-Q-35 LTT 7987 was used,
for GN-2008B-Q-111 G191-B2B was used, and for GS-2010A-Q-58 LTT 3218 was used.
The data were reduced using routines 
supplied in the IRAF Gemini package.

Figure 4 shows the GMOS spectra of the three non-white dwarf objects in our sample, as well as G- and K-type  stellar spectra taken from the spectral atlas of Le Borgne et al. (2003) for reference, for a wavelength range that contains the features useful for classification. The narrow metal lines seen in the spectra exclude the possibility that these are white dwarfs. The three objects (ULAS J1142$+$00, ULAS J1234$+$06N and ULAS J2246$-$00) are early-G to early-K dwarfs or subdwarfs.
For two of these, ULAS J1142$+$00 and ULAS J1234$+$06N, we initially derived too high a proper motion, the revised smaller $H_g$ value places them near the subdwarf region of the RPM (Figure 1). Our true contamination is therefore small, one in fifteen, or 7\%.

Figure 5  shows the spectra for the thirteen newly confirmed white dwarfs. Two of these show weak pressure-broadened H $\alpha$ (ULAS J1323$+$12 and ULAS J2331$+$15), and eleven are featureless.

\subsection{Infrared Photometry}

When carrying out our analysis by fitting models to the observed photometry (\S 4), we found several instances where the infrared photometry appeared to be in error, in particular for the $H$-band. To trace the source of the discrepancies, we obtained repeat photometry for a subset of the sample.

$H$ and $K$ photometry for five of our targets was obtained
using the Wide Field Camera (WFCAM, Casali et al. 2007) on UKIRT, through the UKIRT Service program, via program USERV1876.  
Single-pointing WFCAM ``paw-print'' observations were defined to place the target in one of the four cameras: camera number 2.
Exposure times of 10~s were used, with an 8 position telescope dither pattern, and a 4 position microstep pattern. These 32 exposures were repeated three times, for a total on-source time of 16 minutes, in each filter. An observing log is given in Table 4, together with the derived photometry. 
The ``flat-file'' FITS images and catalogs produced by the CASU pipeline were accessed through the WSA (\S2). The 2$\arcsec$-diameter aperture photometry, provided in counts, were converted to magnitudes using the airmass and zeropoints provided for each of the three reduced groups of observations for each filter and each object. We derived the weighted mean (using the uncertainties provided in the catalog) and the uncertainty in the mean for these measurements. Subsequently, the merged catalog data was released for the three objects observed on 2010 June 13,
and the photometry from this catalog, database USERV1876v20101104, is given in the Table for ULAS J1206$+$03, ULAS J1351$+$12 and ULAS J1404$+$13. The merged catalog photometry agreed with the flatfile photometry to within 0.02 magnitudes for the two brighter sources, within 3\% at $H$ for the fainter ULAS J1404$+$13, and within 12\%
at $K$ for ULAS J1404$+$13, which has $K \approx 19.7$. These differences are well within the quoted uncertainties.

Additional $H$, or $H$ and $K$, photometry was obtained for five more white dwarfs using NIRI (Hodapp et al. 2003) on Gemini North, via program
GN-2011A-Q-59. Exposure times of 15~s or 30~s were used, with a 5 or 9 position telescope dither pattern. The total integration time is given in Table 4, together with the derived photometry. The data were reduced using routines 
supplied in the IRAF Gemini package. UKIRT Faint Standards  were used for calibration (FS 16, FS 33, FS 126, FS 136; Leggett et al. 2006).

Comparing Table 4 to the LAS photometry in Table 2, shows that typically, for the sources fainter than $H \approx 18$, the LAS $H$-band magnitude is too faint by about $2\sigma$. In selecting faint sources near the LAS detection limit, with blue $J-H$ colors, objects have been scattered into the sample with $H$ magnitudes too faint by twice the estimated uncertainty.
A similar effect is seen in searches for faint brown dwarfs that are also blue in $J - H$ (B. Burningham, private communication, 2010).

\section{Analysis}

\subsection{Model Fitting}

The pure-hydrogen, pure-helium and mixed composition model atmospheres used 
in this analysis are described in detail in Kilic et al. (2010a).
These models are in local thermodynamic equilibrium, they allow
energy transport by convection, and they can be calculated with
arbitrary amounts of hydrogen and helium.  For the sample considered here,
the  mixed composition atmospheres had ratios of He to H by number ranging
from low values of $10^{-2}$ to high values of $10^{10}$.
The method used to fit the photometric data is described
at length in Bergeron et al. (2001). Briefly, we convert the magnitudes into observed fluxes using the method of Holberg \& Bergeron (2006) and the appropriate filters. 
We use the SDSS to AB system corrections for $u$, $i$ and $z$ given in 
Eisenstein et al. (2006, Section 2).
Then we fit the resulting energy distributions with those derived from model atmosphere calculations, using a nonlinear least-squares method. Only $T_{\rm eff}$ and the solid angle $\pi(R/D)^2$, where $R$ is the radius of the star and $D$ is its distance from Earth, are considered free parameters. Since no parallax measurements are available, we assume a surface gravity of $\log g = 8$ which determines the value of $R$ for a given value of $T_{\rm eff}$ (Bergeron et al. 2001).  White dwarfs have been
shown to have a very strongly peaked mass and surface gravity
distribution (e.g.\ Bergeron et al.\ 1992; Liebert, Bergeron \& Holberg 2005;
Kepler et al. 2007). DA white dwarfs have a mean mass of 
$0.6 \pm 0.1\ M_{\odot}$ while DBs are slightly more massive with
$0.7 \pm 0.1\ M_{\odot}$; these ranges infer a likely range in gravity
for our sample of $7.7 \leq \log g \leq 8.3$. The photometric variance uncertainties in $T_{\rm eff}$ and the solid angle are
obtained directly from the covariance matrix of the fit.

Since our models do not include the red wing opacity from Ly~$\alpha$
calculated by Kowalski \& Saumon (2006), we 
neglect here the $u$ bandpass in our fitting procedure
as well as the $g$ bandpass for white dwarfs cooler than 4600~K in hydrogen-rich solutions
(as the missing Ly~$\alpha$ opacity has a larger impact at lower temperatures). 
The Ly~$\alpha$ opacity affects a wavelength region where there is very little flux; 
hence the atmospheric structure is not affected significantly by a change of the opacity in the 
ultraviolet, although the predicted blue colors are.

As described in \S 3.2, we have found that in selecting faint sources near the LAS detection limit, with blue $J-H$ colors, objects have been scattered into the sample with $H$ magnitudes too faint by twice the estimated uncertainty.
In our fits to the data we use the NIRI and WFCAM photometry where available (Table 4), and where LAS photometry is used
we exclude data with uncertainties $\geq$0.15 magnitudes. The fits use the DR8 releases of both the SDSS and LAS photometry, the most recent at the time of writing.

We use the energy distributions together with the optical spectra at H~$\alpha$ to constrain the surface composition. 
Only two of the white dwarfs show H~$\alpha$ (Figure 5), however this does not necessarily imply that
the remaining objects are helium-rich, as H~$\alpha$ absorption is not seen in hydrogen-rich white dwarfs
cooler than $\sim 5000$~K.
Figures 6 though 11 show the model fits to the observational data, assuming
 $\log g = 8.0$. 

The energy distributions are somewhat sensitive to $\log g$ in the temperature range considered
here. A variation of $\pm0.3$ dex in $\log g$ yields differences in effective temperature of between 40~K and 80~K.
The uncertainty due to the photometric variations is between 40~K and 150~K. 
Another error estimate is provided by fitting the observed H~$\alpha$ line profile for the two DAs:
the derived $T_{\rm eff}$ differs from that derived from the energy distribution by 100 -- 200 K (for ULAS J1323$+$12
5400~K cf. 5260~K, for ULAS J2331$+$15 5200~K cf. 4970~K).
This is consistent with our other error estimates.

\subsection{Derived Properties}

Table 5 lists the derived atmospheric properties of the 13 new white dwarfs in our sample, together
with the SDSS white dwarf recovered here. Using the 
composition and temperature, and assuming that these stars have the canonical 
white dwarf mass of 0.6 $M_{\odot}$, we can use the synthetic colors of 
Holberg \& Bergeron (2006, an extension of Bergeron, Wesemael \& Beauchamp 
1995) and the evolutionary sequences of Fontaine, Brassard \& Bergeron (2001) 
to derive both a cooling age and distance, and hence tangential velocity. 
These values are also given in Table 5. The derived tangential velocities, $H_g$ and $g - i$ are consistent with
the 40--150 kms$^{-1}$ region indicated in Figure 1.
The uncertainty in  the implied 
cooling age, distance and velocity are primarily due to the
uncertainty in gravity (or mass).  The sense of the gravity effect is that more massive white dwarfs 
will have a longer cooling age and be closer and slower, and vice versa. 

Four of the thirteen  newly identified white dwarfs ($\sim$ 30\%) are best fit with pure-hydrogen atmospheres (Figures 6 and 7; ULAS J0826$-$00, ULAS J1323$+$12, ULAS J1345$+$15, ULAS J2331$+$15), and four ($\sim$ 30\%) are best fit with pure-helium atmospheres (Figure 8; ULAS J1006$+$09, ULAS J1206$+$03, ULAS J1351$+$12, ULAS J1454$-$01). An additional white dwarf (ULAS J1320$+$08) is fit with either pure-helium or a helium-rich atmosphere, where the latter fit, with H/He $= 10^{-4.4}$, is superior (Figure 9). Two sources  (ULAS J0840$+$05, ULAS J1404$+$13) have mixed atmospheres with large flux deficits in the infared, due to pressure-induced H$_2$ absorption (Figure 10). 
As described in Kilic et al. (2010a), there are generally two solutions to such objects. The H$_2$ opacity peaks around 
H/He $= 10^{-2}$ and there is usually a good solution both above and below this peak with  
slightly different temperatures. This is demonstrated by the fits shown in Figure 10, where ULAS J0840$+$05 is fit with 
H/He $= 10^{-4.5}$ or $10^{-0.6}$, and ULAS J1404$+$13 is fit with H/He $= 10^{-2.8}$ or $10^{-2.1}$. 
Thus three white dwarfs ($\sim$ 25\%) have mixed atmospheres, and are similar to the mixed composition white dwarfs found by Kilic et al. (2010a, their Figure 14). There are seven white dwarfs ($\sim$ 55\%) that have pure-helium or helium-dominated atmospheres.
Finally, there are two white dwarfs ($\sim$ 15\% of the sample)
for which we cannot constrain the atmospheric composition:  ULAS J0121$-$00 and ULAS J1436$+$05 (Figure 11). ULAS J1436$+$05 is particularly puzzling --- it appears to be much too bright at $zYJ$. A binary solution does not seem likely given the flux distribution, and the photometric uncertainties are small. 

\subsection{Discussion}

Typically, studies of samples of cool white dwarfs find a much larger fraction of pure-hydrogen atmospheres:
50--60\%  (Bergeron et al. 2001, Kilic et al. 2010a). However Kilic et al. (2010a) find that $\sim$ 60\%  of their sample of white dwarfs (25 of 40) in the temperature range 4500--5000 K have pure-helium atmospheres while only $\sim$ 20\% have pure-hydrogen,  consistent with our findings. As Kilic et al. state, further work is required to understand if the observed overabundance of helium-rich atmosphere white dwarfs at this temperature range is real, or if Ly~$\alpha$ or H$_2$ opacity
uncertainties or photometric errors have biassed the results.

Seven white dwarfs, half of our sample, have 4120~K $\leq T_{\rm eff} \leq$ 4480~K and cooling ages between 7.3 Gyr and 8.7 Gyr. Their distances are 60--80 pc, and their tangential velocities are 40--85 kms$^{-1}$.  
The main-sequence lifetime 
of these remnants is likely to have been $\sim$2 Gyr (e.g. Catal\'{a}n et al. 2008), hence these white dwarfs are most likely
to be thick disk 10--11 Gyr-old objects. 
Of these seven, we could not constrain the atmospheric composition for two, two have mixed atmospheres, two pure hydrogen and one pure helium.   We note that there are no pure helium atmosphere white dwarfs cooler than 4400 K in the 
Kilic et al. (2010a)  sample. The uncertainty in the photometry would allow ULAS J1006$+$09 to be hydrogen-rich, in which case it is slightly cooler, with $T_{\rm eff} = 4230$ K, and older, with a cooling age of 8.4 Gyr.
 
Four of the white dwarfs have 4610~K $\leq T_{\rm eff} \leq$ 4810~K and cooling ages between 6.5 Gyr and 6.9 Gyr. Their distances are 65--100 pc, and velocities are 60--95 kms$^{-1}$.  The remaining three white dwarfs have  4970~K $\leq T_{\rm eff} \leq$ 5260~K and cooling ages between 4.3 Gyr and 5.8 Gyr. Their distances are 60--100 pc, and velocities are 70--100 kms$^{-1}$.
This half of our sample may consist of thin disk remnants with
unusually high velocities (e.g. Bergeron 2003, Tetzlaff et al. 2011), or lower-mass remnants of 
thick disk or halo late-F or G stars.

\section{Conclusions}

Our search of the 1400 deg$^2$ of SDSS and UKIDSS LAS (Data Release 6) sky for white dwarfs cooler than 5000~K
resulted in 17 candidates.
One of these was a previously known white dwarf, and three proved to be subdwarf or dwarf G- to K-type stars,
as determined by GMOS spectroscopy obtained at Gemini Observatory. Of the thirteen newly confirmed
white dwarfs, two show H~$\alpha$, the remaining are featureless.  The relatively high number of subdwarfs in the sample
is due to errors in the initial proper motion derivation, in one case due to an error in epoch determination, and in another to the binary nature of the source.

We fit the SDSS and LAS $ugrizYJHK$ photometry using model atmospheres which can be calculated with
arbitrary amounts of hydrogen and helium.  
As parallaxes are not known for these objects, we adopted the
canonical white dwarf surface gravity of $\log g = 8.0$, although we explore the impact of varying gravity.
We have found that in searching for objects faint at $H$, objects are scattered into our sample
where the photometry is in error by twice the estimated uncertainty. Repeat $H$- and $K$-band photometry
proved to be necessary to fit the white dwarf energy distributions, where $H$, $K >$ 18 magnitudes. 

Seven of the newly identified white dwarfs have 4120~K $\leq T_{\rm eff} \leq$ 4480~K, and cooling ages between 7.3 Gyr and 8.7 Gyr; these are likely to be
thick disk 10--11 Gyr-old objects.
The other half of the sample has 4610~K $\leq T_{\rm eff} \leq$ 5260~K, and cooling age between 4.3 Gyr and 6.9 Gyr. 
These are either thin disk remnants with
unusually high velocities (e.g. Bergeron 2003, Tetzlaff et al. 2011), or lower-mass remnants of 
thick disk or halo late-F or G stars.  Our earlier search of 280 deg$^2$ of LAS sky (Lodieu et al. 2009) resulted in seven spectroscopically confirmed new white dwarfs, with effective temperature  5400~K $\leq T_{\rm eff} \leq$ 6600~K and
cooling age 1.8--3.6 Gyr;
these are a more extreme example of apparently young objects with high tangential velocities. 

We have determined that pairing the SDSS and LAS databases and using
proper motion and color selections of:
$ H_g > 20.5$, $r < 20.7$, $14 \leq J \leq 19.6$, $H < 18.9$, $0.9 \leq g-r \leq 1.6$, $0.2 \leq r-i \leq 0.6$, $0.6 \leq i-J \leq 1.4$, 
$J-H +(\sqrt{err(J)^2 + err(H)^2}) \leq 0.2$
produces  a reasonably complete sample of 4200~K white dwarfs, which have cooling ages of around 8~Gyr.
When complete, the LAS should provide a sample of about twenty cool, thick disk, white dwarfs, and possibly identify
white dwarf remnants of the halo.
The results presented here for various color selections will be a useful guide for identifying cool white dwarfs in ongoing
near-infrared sky surveys such as UKIDSS and the Visible and Infrared Survey Telescope for Astronomy (VISTA; 
McPherson et al. 2006). This is particularly true for white dwarfs showing significant pressure-induced H$_2$ opacity,
due to either extremely low temperatures or higher-pressure mixed H-He atmospheres. The near-infrared surveys will
be useful for studying the complex spectral evolution of white dwarfs, by providing the ratio
of hydrogen- to helium-rich white dwarfs at different values of $T_{\rm eff}$.



\acknowledgments

We are very grateful to the referee, M. Kilic, for a review that greatly improved the paper. 

Some of the data reported here were obtained as part of the United Kingdom Infrared Telescope
(UKIRT) Service Programme; UKIRT is operated by the Joint Astronomy Centre on behalf of the Science and Technology Facilities Council of the U.K.. This paper makes extensive use of the UKIRT Infrared Deep Sky Survey (UKIDSS);
we are grateful to the WFCAM instrument team, the UKIRT staff, the UKIDSS team, the CASU data processing team, and the WSA group at Edinburgh, all of whom have contributed to the success of the survey.

This paper also makes use of Sloan Digital Sky Survey (SDSS) data. Funding for the SDSS and SDSS-II has been provided by the Alfred P. Sloan Foundation, the Participating Institutions, the National Science Foundation, the U.S. Department of Energy, the National Aeronautics and Space Administration, the Japanese Monbukagakusho, the Max Planck Society, and the Higher Education Funding Council for England. The SDSS Web Site is http://www.sdss.org/. The SDSS is managed by the Astrophysical Research Consortium for the Participating Institutions. The Participating Institutions are the American Museum of Natural History, Astrophysical Institute Potsdam, University of Basel, University of Cambridge, Case Western Reserve University, University of Chicago, Drexel University, Fermilab, the Institute for Advanced Study, the Japan Participation Group, Johns Hopkins University, the Joint Institute for Nuclear Astrophysics, the Kavli Institute for Particle Astrophysics and Cosmology, the Korean Scientist Group, the Chinese Academy of Sciences (LAMOST), Los Alamos National Laboratory, the Max-Planck-Institute for Astronomy (MPIA), the Max-Planck-Institute for Astrophysics (MPA), New Mexico State University, Ohio State University, University of Pittsburgh, University of Portsmouth, Princeton University, the United States Naval Observatory, and the University of Washington.

The Digitized Sky Surveys were produced at the Space Telescope Science Institute under U.S. Government grant NAG W-2166. The images of these surveys are based on photographic data obtained using the Oschin Schmidt Telescope on Palomar Mountain and the UK Schmidt Telescope. This research has made use of the VizieR catalogue access tool, CDS, Strasbourg, France.

This work is based on observations obtained at the Gemini Observatory, which is operated by the 
Association of Universities for Research in Astronomy, Inc., under a cooperative agreement 
with the NSF on behalf of the Gemini partnership: the National Science Foundation (United 
States), the Science and Technology Facilities Council (United Kingdom), the 
National Research Council (Canada), CONICYT (Chile), the Australian Research Council (Australia), 
Minist\'{e}rio da Ci\^{e}ncia e Tecnologia (Brazil) 
and Ministerio de Ciencia, Tecnolog\'{i}a e Innovaci\'{o}n Productiva (Argentina).

SKL and AN's research is supported by Gemini Observatory.
NL was funded by the Ram\'on y Cajal fellowship number 08-303-01-02.
This work is also supported in part by the NSERC Canada and by the Fund
FQRNT (Qu\'ebec).




\clearpage


\begin{figure}
\includegraphics[angle=0,scale=.5]{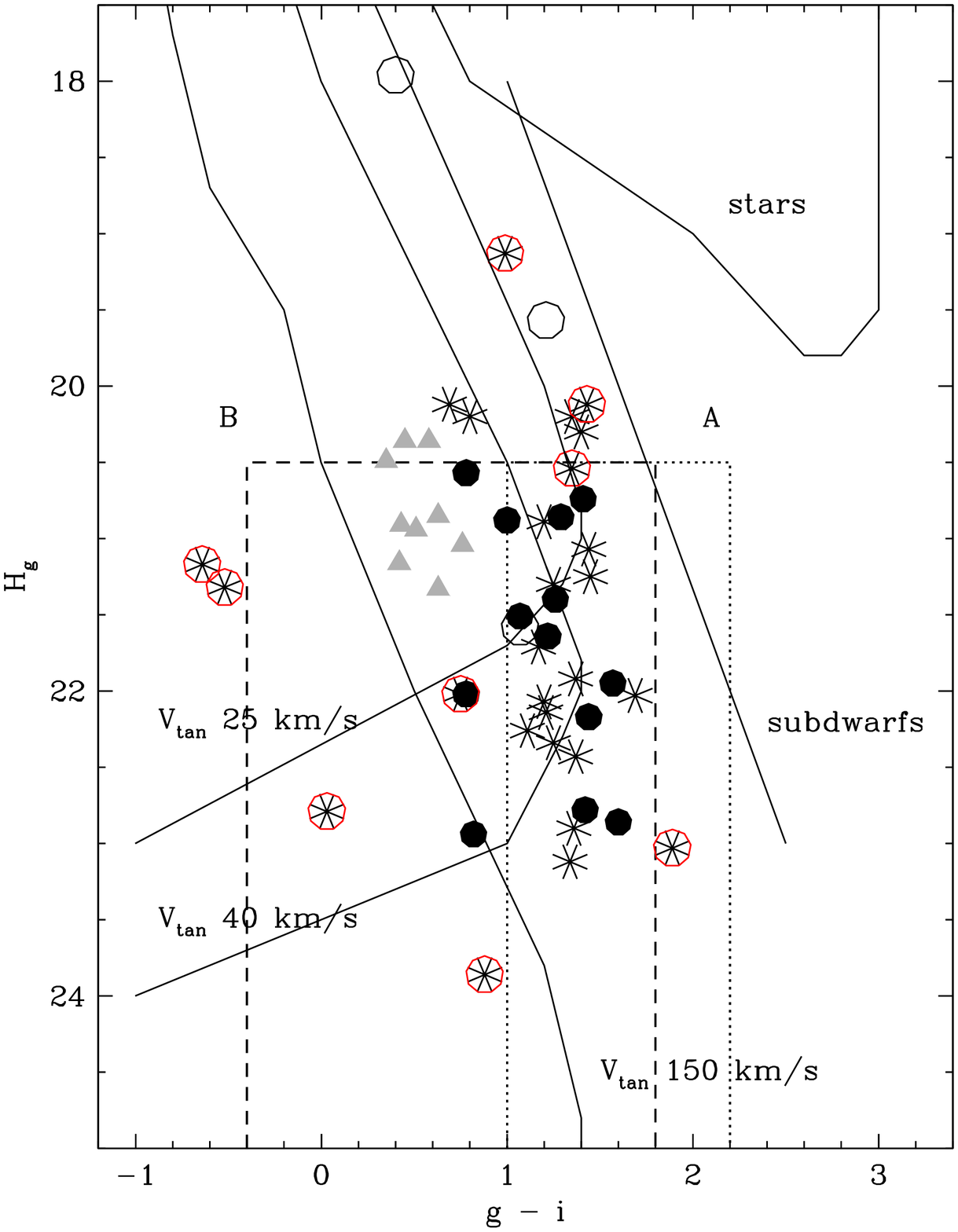}
\caption{Reduced $g$-band proper motion as a function of $g - i$. White dwarf cooling curves for different tangential velocities are shown; the $v_{tan}=$25-40 kms$^{-1}$ curves represent the expected location of disk white dwarfs, and the $v_{\rm tan}=$150 kms$^{-1}$ curve the halo white dwarfs. The typical location of subdwarfs and stars are also shown (based on Figure 6 of Kilic et al. 2010a). Filled circles are confirmed white dwarfs and open circles are G- to K-type (sub)dwarfs  identified in this work. Grey triangles are the warmer sample of confirmed white dwarfs from Lodieu et al. (2009b). Asterisks are cool white dwarfs from Kilic et al. (2010a; white dwarfs identified both by the RPM method and spectroscopically by Gates et al. (2004) and Harris et al. (2008) are shown). Points circled in red are white dwarfs cooler than 4000~K. The regions labelled A and B indicate the search criteria used here, see Figure 2 and the text. The photometry is taken from DR8 of the SDSS, and the proper motions are derived from the DR8 releases of the SDSS and LAS.
\label{fig1}}
\end{figure}

\clearpage 
\begin{figure}
\includegraphics[angle=0,scale=.6]{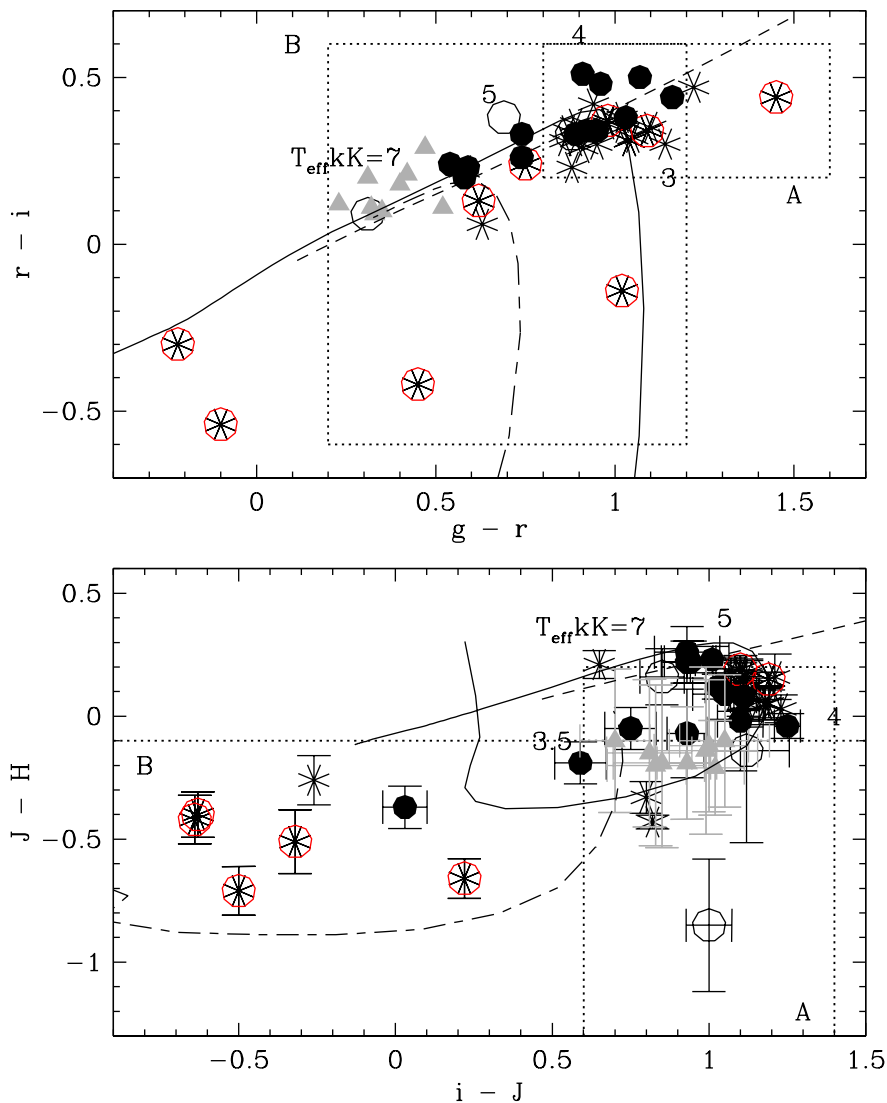}
\caption{Plots demonstrating the color selection of the white dwarf candidates. Model sequences for white dwarfs with pure hydrogen and pure helium atmospheres are shown as solid and dashed lines, and a sequence with He/H=100 is shown as a short-long-dash line. $T_{\rm eff}/1000$~K is indicated along the sequences for the pure hydrogen case. Symbols are as in Figure 1. The regions labelled A and B indicate the search criteria used here, as described in the text. The optical photometry is taken from SDSS DR8, and the infrared from our WFCAM and NIRI observations where available, otherwise from DR8 of the LAS. 
\label{fig2}}
\end{figure}

\clearpage 

\begin{figure}
\includegraphics[angle=0,scale=.7]{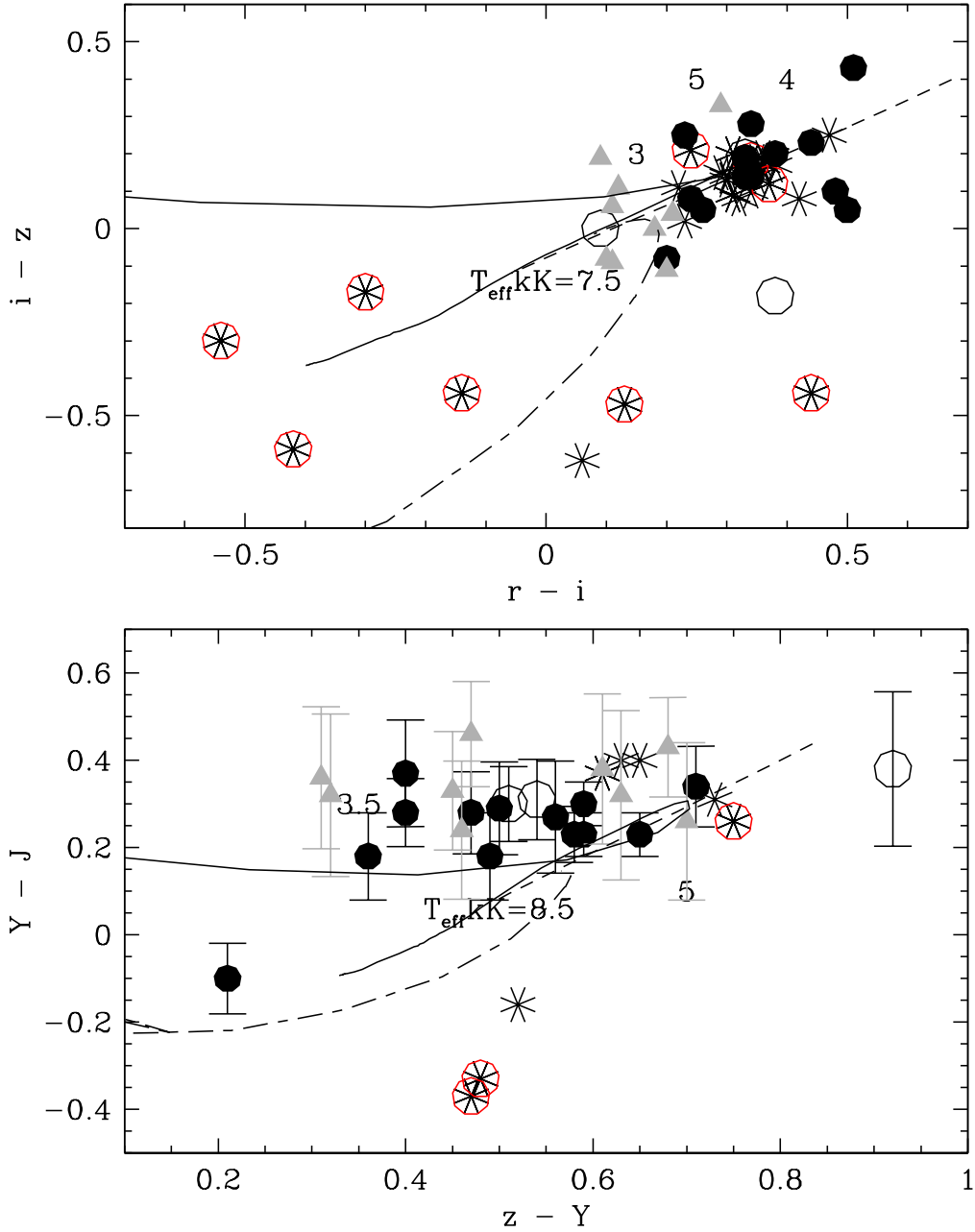}
\caption{Plots of the red to far-red colors of the samples considered here. Symbols, sequences and data sources are as in Figures 2 and 3.
\label{fig3}}
\end{figure}

\clearpage

\begin{figure}
\includegraphics[angle=-90,scale=.6]{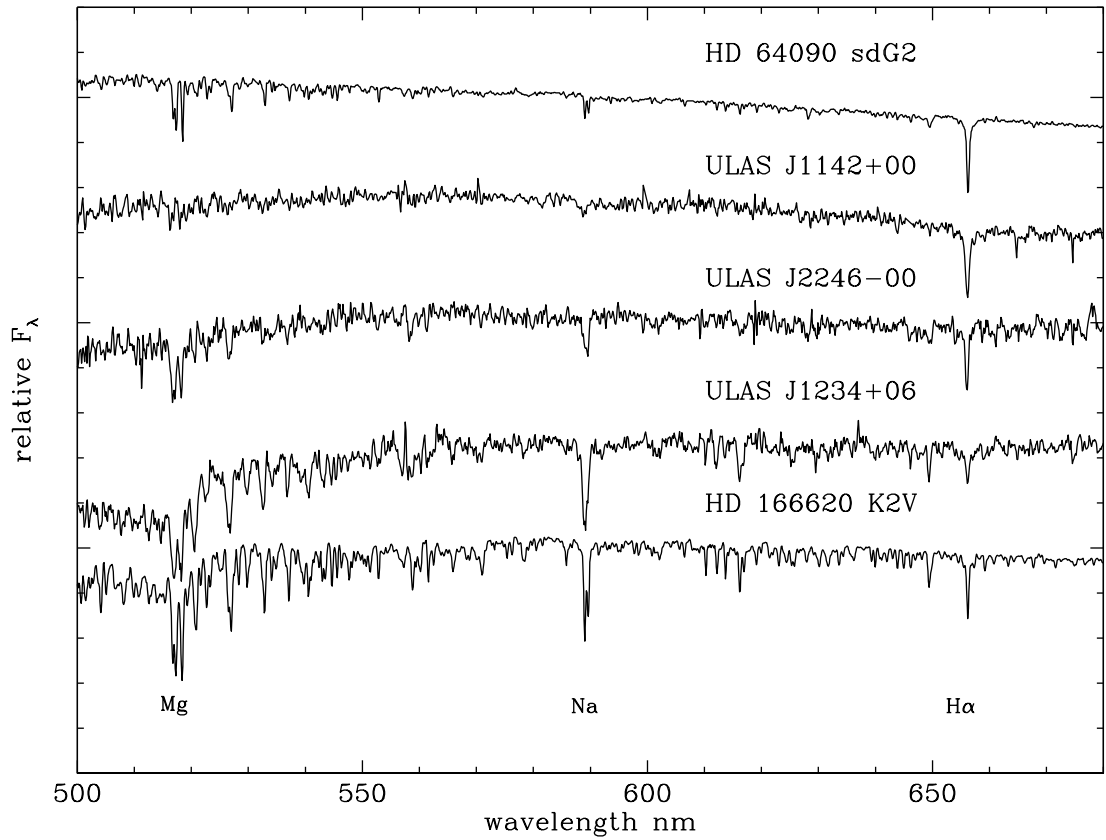}
\caption{GMOS spectra of the three  ULAS G and K stars in our sample, compared to similar spectra taken from Le Borgne et al. (2003; smoothed to the resolution of our data). The spectra have been normalized to unity at 600~nm, and offset by 0.5 flux units for clarity.  
\label{fig4}}
\end{figure}

\clearpage 

\begin{figure}
\includegraphics[angle=-90,scale=.6]{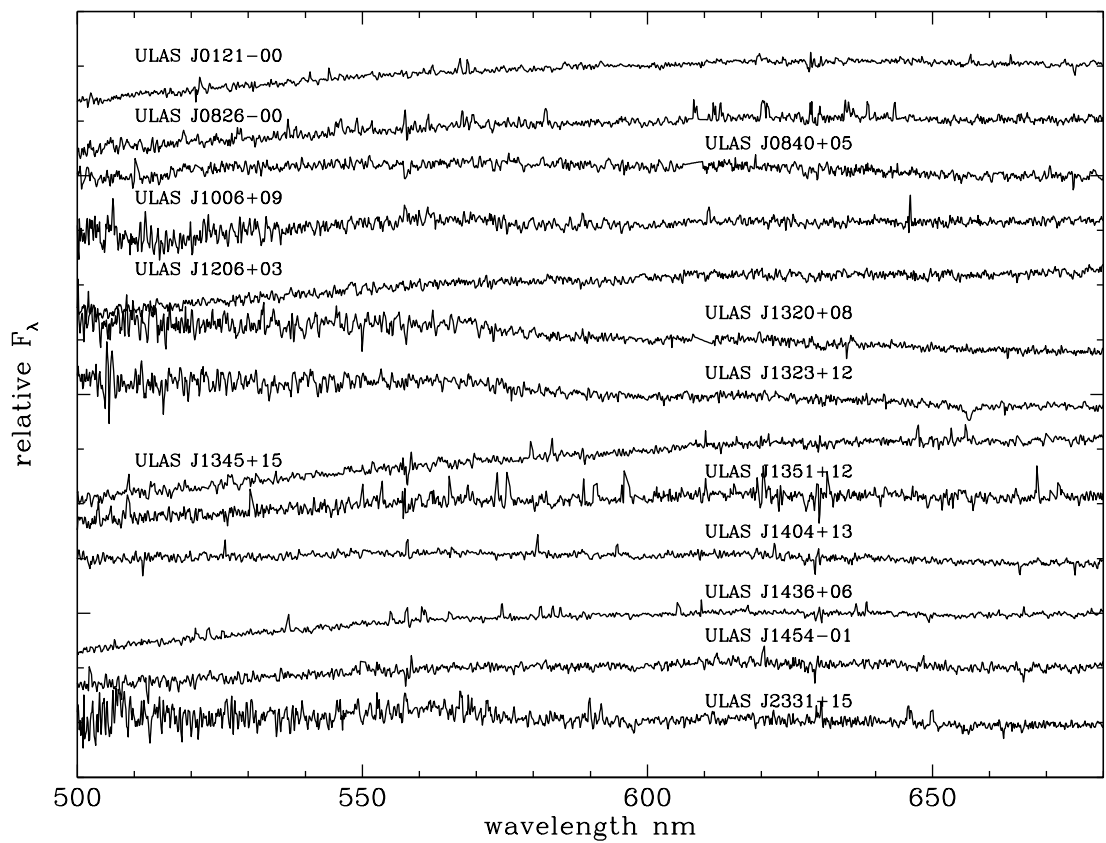}
\caption{GMOS spectra of previously unknown white dwarfs in our sample. The spectra have been normalized to unity at 600~nm, and offset by 0.5 flux units for clarity.  H~$\alpha$ absorption is seen in the ULAS J1323$+$12 spectrum, and weakly in the ULAS J2331$+$15 spectrum (see also Figures 6 and 7).
\label{fig5}}
\end{figure}

\clearpage

\begin{figure}
\includegraphics[angle=0,scale=.8]{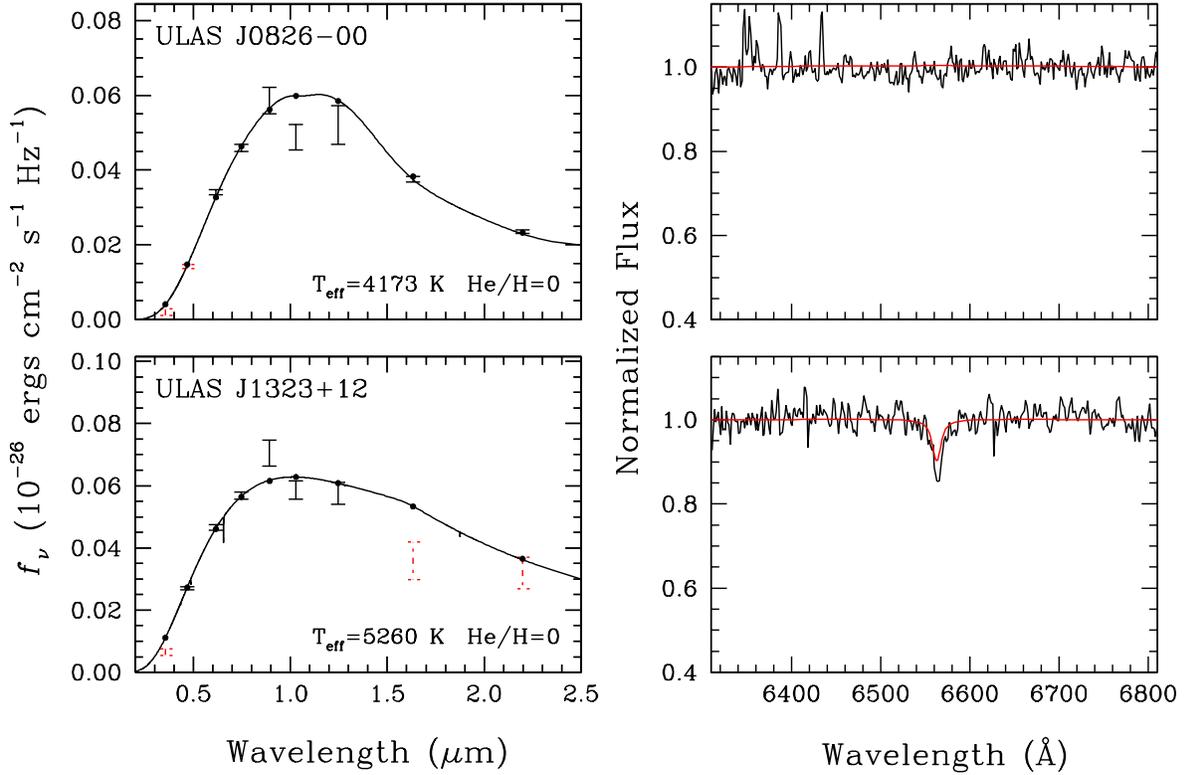}
\caption{Two of the four white dwarfs in our sample best fit with pure hydrogen atmospheres. The error bars in the left panels represent  SDSS $ugriz$ and NIRI, WFCAM and/or LAS $YJHK$ photometry. The dashed error bars indicate photometric datapoints that have been ignored in the fits: $u$ data, as well as $g$ for ULAS J0826$-$00, and $H$ and $K$ for ULAS J1323$+$12 (see text).
The solid lines in the left panels represent the
monochromatic model fluxes while the solid dots correspond to the average
over the filter bandpass.
A surface gravity $\log g=8.0$ is
assumed, and the derived $T_{\rm eff}$  is shown in the legends.
The right panels show the observed spectrum around 
H~$\alpha$, with the modelled pure-hydrogen atmosphere line 
profiles. ULAS J0826$-$00 is too cool to show H~$\alpha$.
\label{fig8}}
\end{figure}

\clearpage 
\begin{figure}
\includegraphics[angle=0,scale=.8]{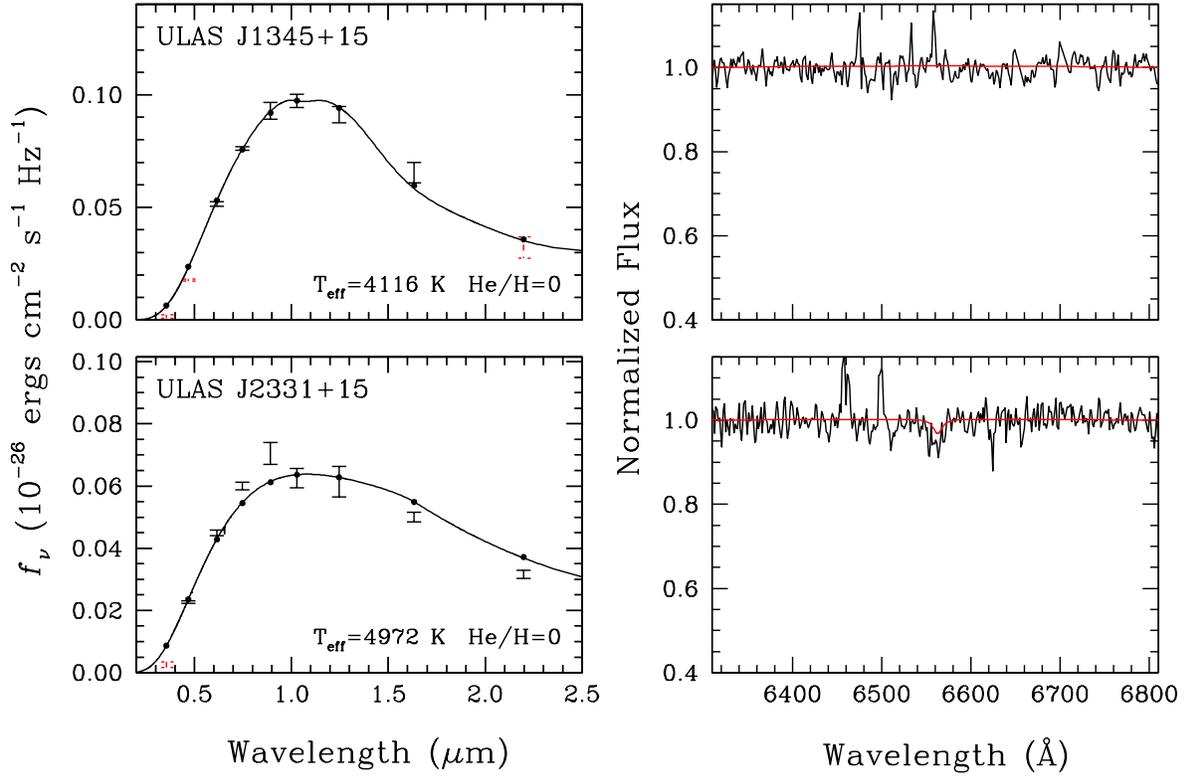}
\caption{The two other white dwarfs in our sample best fit with pure hydrogen atmospheres.
Symbols are as in Figure 6. $u$ data have been ignored in the fits, as well as $g$ and $K$ for
ULAS J1345$+$15 (see text). ULAS J1345$+$15 is too cool to show H~$\alpha$ (right panels).}
\end{figure}

\clearpage 
\begin{figure}
\includegraphics[angle=0,scale=.8]{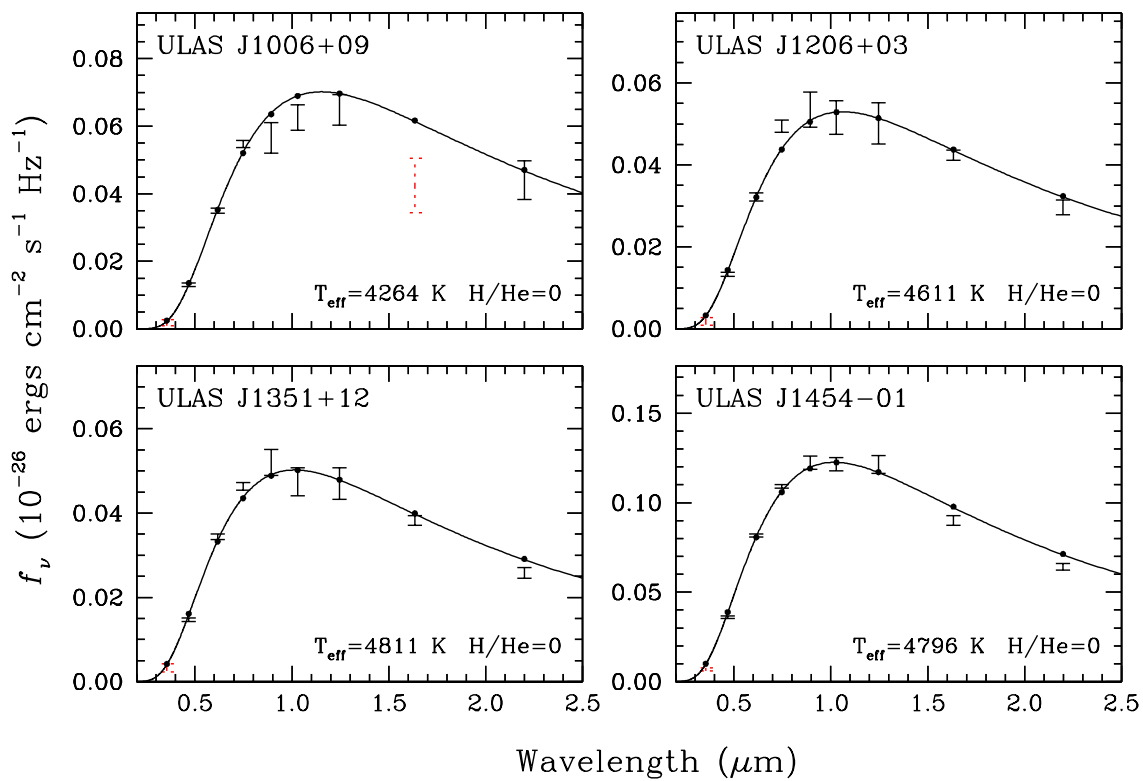}
\caption{Four white dwarfs in our sample best fit with pure helium atmospheres. Symbols are as in Figure 6. $u$ data have been ignored in the fits, as well as
$H$ for ULAS J1006$+$09 (see text).
All objects have observed and modelled featureless spectra (not shown).
\label{fig9}}
\end{figure}

\clearpage 

\begin{figure}
\includegraphics[angle=0,scale=.8]{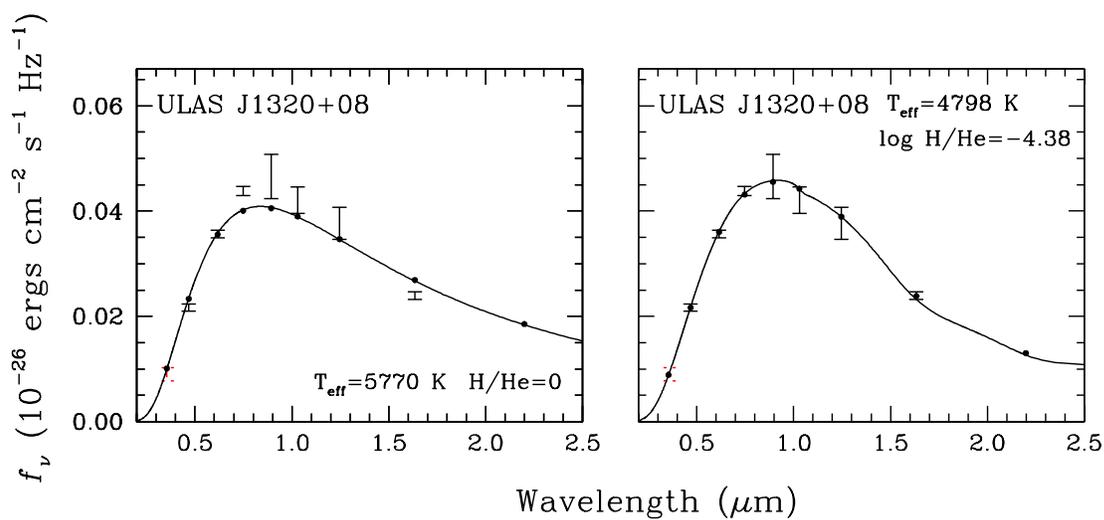}
\caption{One white dwarf in our sample best fit with either a pure helium atmosphere (left panel) or a mixed atmosphere with a high He/H ratio (right panel); the mixed fit is preferred. Symbols are as in Figure 6. $u$ data have been ignored in the fits (see text). Both the observed and modelled spectrum (not shown) is featureless.
\label{fig10}}
\end{figure}

\clearpage 

\begin{figure}
\includegraphics[angle=0,scale=.8]{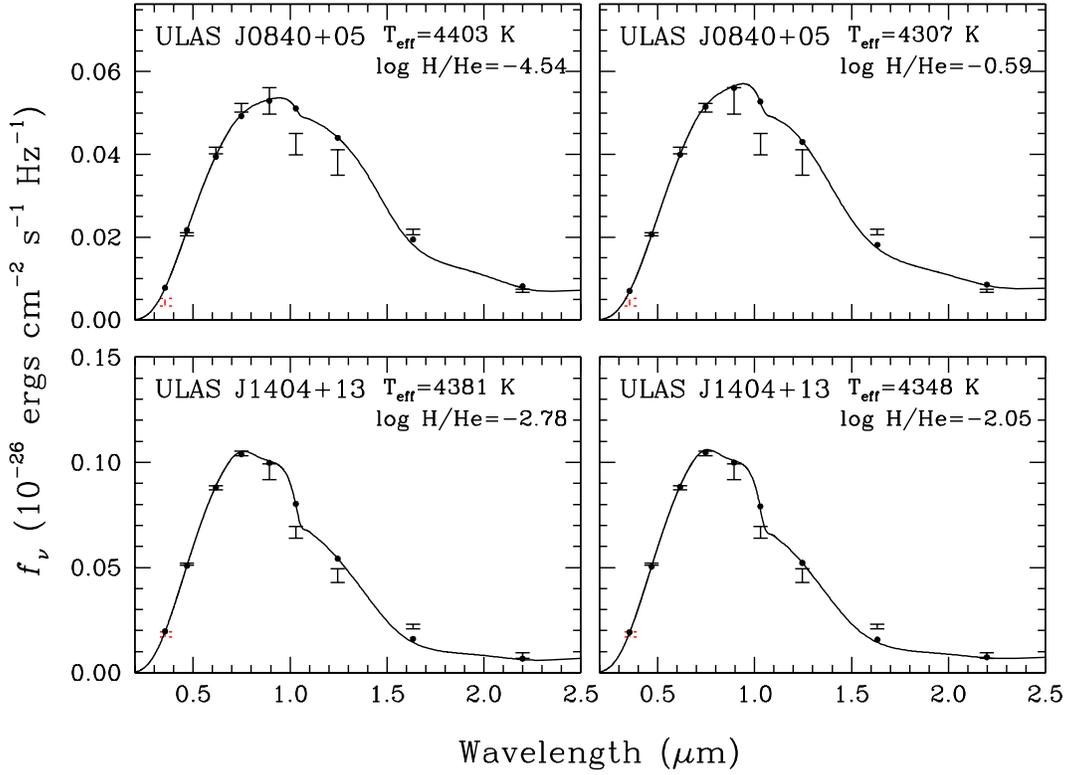}
\caption{Two white dwarfs in our sample best fit with  mixed H/He atmospheres. 
Symbols are as in Figure 6. $u$ data have been ignored in the fits (see text).
Solutions with two values of the He/H ratio are possible (see text).
Both objects have observed and modelled featureless spectra (not shown).
\label{fig11}}
\end{figure}

\clearpage 

\begin{figure}
\includegraphics[angle=0,scale=.8]{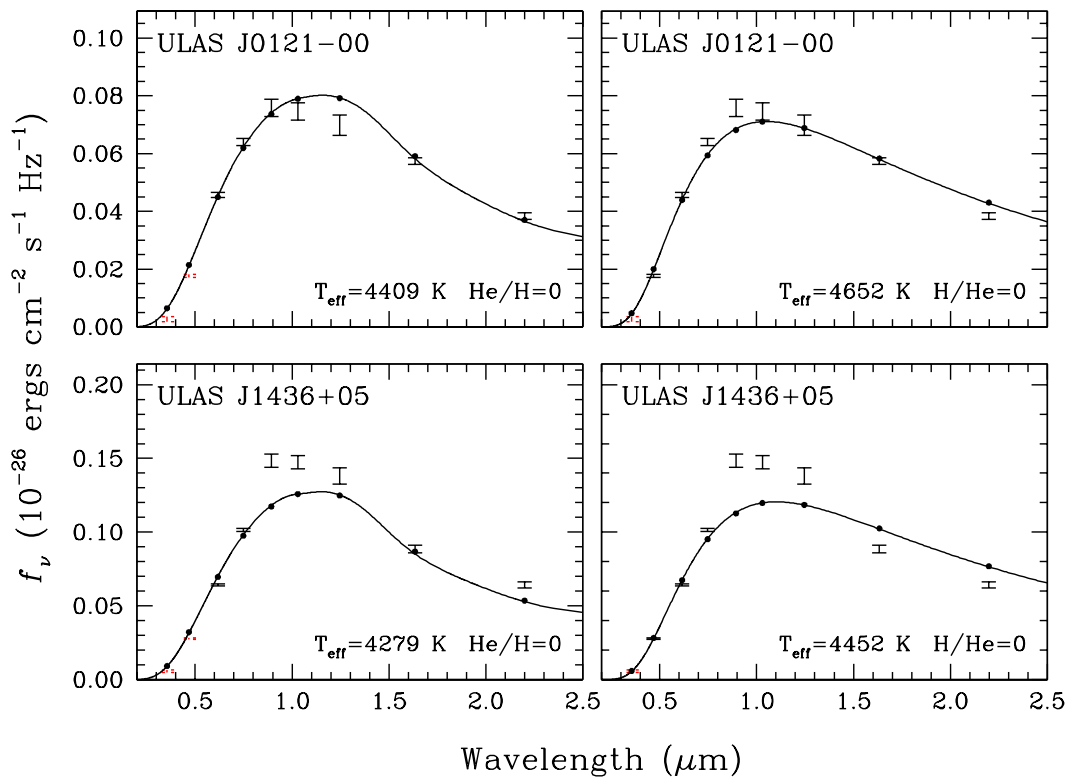}
\caption{Two white dwarfs in our sample for which composition could not be constrained, the pure hydrogen
fit is shown in the left panels, and the pure helium fit in the right panels. Symbols are as in Figure 6. $u$ data have been ignored in the fits, as well as $g$ in the pure hydrogen fits. Both objects have observed and modelled featureless spectra (not shown).}
\end{figure}

\clearpage 





\clearpage

\begin{deluxetable}{lcccccccclc}
\tabletypesize{\scriptsize}
\tablewidth{0pt}
\rotate
\tablecaption{Astrometry for Candidate White Dwarfs}
\tablehead{
\colhead{Short Name} & \colhead{Right Ascension} & \colhead{Declination} & \colhead{Epoch} & 
\multicolumn{2}{c}{SDSS-LAS $\mu$ $\arcsec$yr$^{-1}${\tablenotemark{a}}} & \multicolumn{2}{c}{Other $\mu$ $\arcsec$yr$^{-1}${\tablenotemark{b}}} & \colhead{$\mu${\tablenotemark{b}}}
& \colhead{RPM{\tablenotemark{c}}} & \colhead{Search{\tablenotemark{d}}}\\
\colhead{} & \colhead{HH:MM:SS.SS} & \colhead{DD:MM:SS.S} & \colhead{YYYYMMDD} & \colhead{RA} & \colhead{Dec} 
& \colhead{RA} & \colhead{Dec}
& \colhead{Note}  & \colhead{$H_g$} & \colhead{Region}\\
}
\startdata
ULAS J0121$-$00{\tablenotemark{e}} & 01:21:03.02 & $-$00:38:33.4 & 20061221  & $+$0.068 & $+$0.071 
& $+$0.125 & $+$0.044 & 1 & 20.74 [21.39] & A\\
ULAS J0826$-$00{\tablenotemark{e}} &  08:26:44.57 & $-$00:35:47.6 & 20061205 & $+$0.087  & $+$0.032 
& $+$0.126  & $+$0.033 & 2  & 20.86 [21.59] & A \\
ULAS J0840$+$05{\tablenotemark{e}} & 08:40:01.43 & $+$05:15:28.8 & 20061216 & $-$0.030  &$-$0.109  
& -0.102 & -0.076 & 2 & 20.88 [21.13] & B \\
ULAS J1006$+$09{\tablenotemark{e}} & 10:06:05.14 & $+$09:50:15.2 & 20070421  & $+$0.131 & $-$0.067 
&$+$0.134  & $-$0.053 & 1 & 21.95 [21.90] & A \\
ULAS J1142$+$00{\tablenotemark{f}} & 11:42:48.61 & $+$00:12:26.0 & 20080415 & $+$0.050 & $+$0.011 
& $-$0.007& $-$0.007 & 1 & 17.96 [14.39]{\tablenotemark{g}} & C\\
ULAS J1206$+$03{\tablenotemark{e}} & 12:06:05.93 & $+$03:47:17.1 &  20070405  & $+$0.021 & $-$0.163 
& $-$0.008 & $-$0.160 & 2 & 22.17 [22.11] & B \\
ULAS J1234$+$06N & 12:34:40.82 & $+$06:47:19.8  & 20070121 &  $+$0.048 & $-$0.018 
& & & & 19.57{\tablenotemark{h}} & B \\
ULAS J1234$+$06S{\tablenotemark{i}} & 12:34:40.79 & $+$06:47:18.6  & 20070121 &  $+$0.072 & $-$0.001 
& & &  & 21.72{\tablenotemark{h}} &  \\
SDSS J1247$+$06{\tablenotemark{j}} & 12:47:38.95 & $+$06:46:04.8 &  20070121 & $-$0.383 & $+$0.072 
& $-$0.385 & $+$0.071 & 1 & 22.98 [22.99]  & A\\
ULAS J1320$+$08{\tablenotemark{e}} & 13:20:16.50 & $+$08:36:43.6 &   20070220 & $-$0.039 & $-$0.192 
& $+$0.019 & $-$0.208 & 1 & 22.02 [21.62]  & B\\
ULAS J1323$+$12{\tablenotemark{e}} & 13:23:57.27 & $+$12:03:13.4  & 20070417  & $+$0.303 & $-$0.140 
& $+$0.120 & $-$0.171 & 3 & 22.94 [21.49]  & Ab\\
ULAS J1345$+$15{\tablenotemark{e}} & 13:45:50.68 & $+$15:14:54.5  &  20080601 &  $-$0.139 & $-$0.221
& $-$0.076 & $-$0.246 & 3 & 22.86 [22.83] & A\\
ULAS J1351$+$12{\tablenotemark{e}} & 13:51:09.60 & $+$12:14:08.5  &  20070421  & $+$0.024 & $-$0.119
& $+$0.002 & $-$0.104 & 1 & 21.40 [21.07] & A\\
ULAS J1404$+$13{\tablenotemark{e}} & 14:04:51.86 & $+$13:30:55.6  & 20080311  & $-$0.106 & $+$0.113 
& $-$0.099 & $+$0.106 & 1 & 20.57 [20.43] & B \\
ULAS J1436$+$05{\tablenotemark{e}} & 14:36:02.52 & $+$05:38:01.0  & 20080526  & $-$0.092 & $-$0.301 
& $-$0.102 & $-$0.284 & 2 & 22.78 [22.69] & A\\
ULAS J1454$-$01{\tablenotemark{e}} & 14:54:27.16 & $-$01:10:05.8 & 20080417  & $-$0.209 & $-$0.035 
& $-$0.263 & $-$0.059 & 2 & 21.64 [22.16] & A\\
ULAS J2246$-$00{\tablenotemark{f}} & 22:46:29.89 & $-$00:50:53.1 & 20050906  &  $-$0.111 & $-$0.009 
& $-$0.097 & $+$0.012 & 3 & 21.58 [21.30]  & B\\
ULAS J2331$+$15{\tablenotemark{e}} & 23:31:39.91 &  $+$15:18:00.4 &  20070930 & $+$0.136 & $-$0.082 
& $+$0.113 & $-$0.097 & 1 & 21.51 [21.37] & A \\
\enddata
\tablenotetext{a}{Uncertainty in proper motion is $\sim 14$ mas yr$^{-1}$, except for ULAS J1323$+$12 for which the uncertainty is 30 mas yr$^{-1}$.}
\tablenotetext{b}{Other sources: (1) SDSS-USNO proper motion reported in DR8 of the SDSS; (2) derived here from digital sky images together with the SDSS and LAS astrometry; (3) derived here from digital sky images, as well as the USNO-B, SDSS and LAS astrometry. Typical uncertainty is $\sim 10$ mas yr$^{-1}$.}
\tablenotetext{c}{Value in brackets is that implied by the alternative proper motion value.}
\tablenotetext{d}{All region selections include $r_{AB} < 20.7$ and $14 \leq J \leq 19.6$; in addition
for region A  $H_g > 20.5$, $ H < 18.9$, $0.8 \leq g-r \leq 1.6$, $0.2 \leq r-i \leq 0.6$, $0.6 \leq i-J \leq 1.4$, $J-H \leq 0.2$. Region Ab is the same, except that the source is slightly bluer: $g-r = 0.63$. Region B is defined by
$H_g > 20.5$, $0.2 \leq g-r \leq 1.2$, $-0.6 \leq r-i \leq +0.6$, $J-H \leq -0.1$, $H-K \leq -0.1$.
Region C is defined by $H_g > 22.5$, $-0.5 \leq g-i \leq 1.0$, $0.0 \leq r-i \leq 0.6$, $H < 18.8$, $J-H \leq 0.1$, $H-K \leq 0$.}
\tablenotetext{e}{Confirmed as a white dwarf spectroscopically in this work.}
\tablenotetext{f}{Confirmed as a G- to K-type (sub)dwarf star spectroscopically in this work.}
\tablenotetext{g}{Erroneously high initial proper motion determination placed object in sample.}
\tablenotetext{h}{This close pair was incorrectly matched, resulting in a high initial proper motion determination which placed the Northern object in sample.}
\tablenotetext{i}{No optical spectrum obtained.}
\tablenotetext{j}{Previously discovered in SDSS by Kilic et al. (2010a).}

\end{deluxetable}

\clearpage
%
%
\begin{deluxetable}{lrrrrrrrrr}
\tabletypesize{\scriptsize}
\tablewidth{0pt}
\rotate
\tablecaption{SDSS DR8 and UKIDSS LAS DR8 Photometry for Candidate White Dwarfs}
\tablehead{
\colhead{Short Name} & \colhead{$u$(err)} & \colhead{$g$(err)} & \colhead{$r$(err)} & \colhead{$i$(err)} &
\colhead{$z$(err)}  & \colhead{$Y$(err)} & \colhead{$J$(err)} & \colhead{$H$(err)} & \colhead{$K$(err)}  \\
}
\startdata
ULAS J0121$-$00{\tablenotemark{a,b}} & 22.87(0.30) & 20.78(0.03) & 19.75(0.02) & 19.37(0.02) & 19.17(0.04) & 18.59(0.04) & 18.36(0.05) & 18.56(0.18) & 18.06(0.23)\\
ULAS J0826$-$00{\tablenotemark{a,b}} & 23.19(0.44) & 21.02(0.04) & 20.07(0.02) & 19.73(0.02) & 19.45(0.06) & 19.05(0.07) & 18.68(0.10) & 18.75(0.17) &   \\
ULAS J0840$+$05{\tablenotemark{a,b}} & 22.37(0.21) & 20.61(0.02) & 19.87(0.02) & 19.61(0.02) & 19.56(0.06) & 19.20(0.06) & 19.02(0.08) & 19.21(0.18) &  \\
ULAS J1006$+$09{\tablenotemark{a}} & 23.26(0.48) & 21.11(0.04) & 20.04(0.02) & 19.54(0.02) & 19.49(0.08) & 18.78(0.06) & 18.44(0.07) & 18.46(0.19) & 17.90(0.13) \\
ULAS J1142$+$00{\tablenotemark{c}} & 20.47(0.04) & 19.41(0.01) & 19.10(0.01) & 19.01(0.01) & 19.01(0.03) & 18.47(0.06) & 18.16(0.07) & 18.00(0.09) & 17.95(0.15) \\
ULAS J1206$+$03{\tablenotemark{a,b}} & 23.30(0.52) & 21.09(0.04) & 20.13(0.03) & 19.65(0.03) & 19.55(0.08) & 18.99(0.08) & 18.72(0.10) & 18.98(0.22) & 18.46(0.22) \\
ULAS J1234$+$06N{\tablenotemark{c}} & 22.74(0.31) & 21.02(0.03) & 20.14(0.02) & 19.81(0.02) & 19.62(0.06) & 19.11(0.05) & 18.81(0.07) & 19.66(0.26) & 17.95(0.13)  \\
ULAS J1234$+$06S{\tablenotemark{d}} & 25.37(0.65) & 22.43(0.10) & 21.13(0.05) & 20.57(0.04) & 22.27(0.44)
& 19.61(0.08) & 19.08(0.09) & 19.28(0.18) & \\ 
SDSS J1247$+$06{\tablenotemark{e}} & 20.95(0.08) & 20.03(0.02) & 18.68(0.01) & 18.39(0.01) & 18.27(0.02) & 17.76(0.02) & 17.54(0.03) & 17.49(0.04) & 17.54(0.10)\\
ULAS J1320$+$08{\tablenotemark{a,b}} & 21.56(0.14) & 20.56(0.03) & 20.02(0.02) & 19.78(0.02) & 19.70(0.09) & 19.21(0.06) & 19.03(0.08) & 19.53(0.16) & \\
ULAS J1323$+$12{\tablenotemark{a}} & 21.91(0.17) & 20.32(0.02) & 19.73(0.02) & 19.50(0.02) & 19.25(0.06) & 18.85(0.05) & 18.57(0.06) & 18.64(0.17) & 18.25(0.16) \\
ULAS J1345$+$15{\tablenotemark{a}} & 23.52(0.50) & 20.78(0.03) & 19.61(0.02) & 19.18(0.01) & 18.95(0.04) & 18.30(0.03) & 18.07(0.04) & 17.99(0.07) & 18.24(0.15) \\
ULAS J1351$+$12{\tablenotemark{a,b}} & 22.66(0.29) & 20.98(0.03) & 20.06(0.02) & 19.72(0.02) & 19.58(0.06)  & 19.08(0.07) & 18.79(0.08) & 18.88(0.16) & 18.63(0.19) \\
ULAS J1404$+$13{\tablenotemark{a,b}} & 20.79(0.07) & 19.62(0.01) & 19.04(0.01) & 18.84(0.01) & 18.92(0.04)  & 18.71(0.04) & 18.81(0.07) & 19.01(0.15) & \\
ULAS J1436$+$05{\tablenotemark{a,b}} & 22.05(0.17) & 20.29(0.02) & 19.38(0.01) & 18.87(0.01) & 18.44(0.03)  & 17.85(0.03) & 17.62(0.04) & 17.51(0.06) & 17.47(0.11)\\
ULAS J1454$-$01{\tablenotemark{a,b}} & 21.86(0.12) & 20.01(0.02) & 19.12(0.01) & 18.79(0.01) & 18.65(0.03)  & 18.06(0.03) & 17.76(0.04) & 17.68(0.09) & 17.56(0.12)\\
ULAS J2246$-$00{\tablenotemark{c}} & 23.28(0.53) & 21.35(0.05) & 20.66(0.03) & 20.28(0.04) & 20.46(0.16)  & 19.54(0.12) & 19.16(0.13) & 19.30(0.35) &\\
ULAS J2331$+$15{\tablenotemark{a,b}} & 22.92(0.33) & 20.51(0.02) & 19.77(0.02) & 19.44(0.02) & 19.25(0.05) & 18.78(0.05) & 18.50(0.08) & 18.72(0.25) &\\
\enddata
\tablecomments{SDSS $ugriz$ magnitudes are on the AB system (Fukugita et al. \ 1996). LAS $YJHK$ are on the Vega
system (Hewett et al.\ 2006).}
\tablenotetext{a}{Confirmed as a white dwarf spectroscopically in this work.}
\tablenotetext{b}{Improved $H$ and $K$ photometry is given in Table 4.}
\tablenotetext{c}{Confirmed as  a G- to K-type (sub)dwarf star spectroscopically in this work.}
\tablenotetext{d}{No spectrum obtained.}
\tablenotetext{e}{Previously discovered in SDSS by Kilic et al. (2010a).}
\end{deluxetable}

\clearpage

%
%
\begin{deluxetable}{lrccl}
\tabletypesize{\footnotesize}
\tablecaption{GMOS Observation Log}
\tablewidth{0pt}
\tablehead{
\colhead{Short Name} & \colhead{SDSS $r$} & \colhead{Total Exp.} & \colhead{Date} & Program \\
\colhead{}  & \colhead{AB}  &\colhead{seconds} & \colhead{YYYYMMDD} & \\
}
\startdata
ULAS J0121$-$00{\tablenotemark{a}} & 19.70 & 5100 & 20100710 & GS-2010A-Q-58 \\
ULAS J0826$-$00{\tablenotemark{a}} & 20.07 & 8400 & 20100119 & GS-2010A-Q-58 \\
ULAS J0840$+$05{\tablenotemark{a}} & 19.83 & 3600 & 20081209 & GS-2008B-Q-35 \\
ULAS J1006$+$09{\tablenotemark{a}} & 20.04 & 6400 & 20090122 & GN-2008B-Q-111 \\
ULAS J1142$+$00{\tablenotemark{b}} & 19.11 & 1800 & 20100119 & GS-2010A-Q-58 \\
ULAS J1206$+$03{\tablenotemark{a}} & 20.13 & 5040 & 20090120 & GS-2008B-Q-35 \\
ULAS J1234$+$06N{\tablenotemark{b,c}} & 19.76 & 5760 & 20090122 & GS-2008B-Q-35 \\
ULAS J1320$+$08{\tablenotemark{a}} & 20.02 & 6400 & 20090126, 20090127 & GN-2008B-Q-111 \\
ULAS J1323$+$12{\tablenotemark{a}} & 19.77 & 4800 & 20090122, 20090125  & GN-2008B-Q-111\\
ULAS J1345$+$15{\tablenotemark{a}} & 19.61 & 4400 & 20100415  & GS-2010A-Q-58 \\
ULAS J1351$+$12{\tablenotemark{a}} & 20.07 & 8400 & 20100418, 20100607 & GS-2010A-Q-58 \\
ULAS J1404$+$13{\tablenotemark{a}} & 19.04 & 1800 & 20100508 & GS-2010A-Q-58 \\
ULAS J1436$+$05{\tablenotemark{a}} & 19.40 & 3200 & 20100607, 20100713 & GS-2010A-Q-58 \\
ULAS J1454$-$01{\tablenotemark{a}} & 19.12 & 1800 & 20100418 & GN-2008B-Q-111\\
ULAS J2246$-$00{\tablenotemark{b}} & 20.64 & 11400 & 20080731 & GS-2008B-Q-35 \\
ULAS J2331$+$15{\tablenotemark{a}} & 19.77 & 4800 & 20081221  & GN-2008B-Q-111\\
\enddata
\tablenotetext{a}{Confirmed as a white dwarf spectroscopically in this work.}
\tablenotetext{b}{Confirmed as a G- to K-type (sub)dwarf star spectroscopically in this work.}
\tablenotetext{c}{Close pair of objects, spectroscopy is for northern source.}
\label{tab:ObsLog}
\end{deluxetable}

\clearpage

%
%
\begin{deluxetable}{lrrccl}
\tabletypesize{\footnotesize}
\tablewidth{0pt}
\tablecaption{WFCAM and NIRI Photometry for LAS White Dwarfs}
\tablehead{
\colhead{Short Name} &  \colhead{$H$(err)} & \colhead{$K$(err)} & \colhead{Exp. $H$, $K$}   & \colhead{Date} & Program\\
\colhead{} & \colhead{} & \colhead{}  & \colhead{minutes} & \colhead{YYYYMMDD} & \colhead{} \\
}
\startdata
ULAS J0121$-$00 & 18.13(0.02) &  18.05(0.03)  & 16, 16 & 20100814,20100815 & USERV 1876 \\
ULAS J0826$-$00 & 18.59(0.02) &  18.58(0.02)  & 9, 9 & 20110126 & GN-2011A-Q-59 \\
ULAS J0840$+$05 & 19.21(0.03) &  19.89(0.06) & 13.5, 18 & 20110125 & GN-2011A-Q-59 \\
ULAS J1206$+$03 & 18.46(0.03) &  18.33(0.06) & 16, 16 & 20100613 & USERV 1876\\
ULAS J1320$+$08 & 19.08(0.03) &           & 22.5 & 20110126 & GN-2011A-Q-59 \\
ULAS J1351$+$12 & 18.57(0.03) &  18.48(0.05)  & 16, 16 & 20100613 & USERV 1876\\
ULAS J1404$+$13 & 19.18(0.05) &  19.72(0.16) & 16, 16 & 20100613 & USERV 1876\\
ULAS J1436$+$05 & 17.66(0.03) & 17.49(0.03) & 1.25, 1.25 & 20110105 & GN-2011A-Q-59 \\
ULAS J1454$-$01 & 17.64(0.03) & 17.49(0.03) & 1.25, 1.25 & 20110209 & GN-2011A-Q-59 \\
ULAS J2331$+$15 & 18.28(0.03) &  18.26(0.04)  & 16, 16 &20100807, 20100812 & USERV 1876\\

\enddata
\tablecomments{$HK$ are on the Vega
system (Tokunaga \& Vacca 2005, Hewett et al.\ 2006).  }
\end{deluxetable}

\clearpage



%
\begin{deluxetable}{lccccrr}
\tabletypesize{\footnotesize}
\tablewidth{0pt}
\tablecaption{Derived Properties of the LAS White Dwarfs}
\tablehead{
\colhead{Short Name} & Spectral & \colhead{Atmospheric} & \colhead{$T_{\rm eff}${\tablenotemark{a}}}
& \colhead{Cooling{\tablenotemark{b}}} &  \colhead{Distance{\tablenotemark{c}}}
& \colhead{$v_{\rm tan}${\tablenotemark{d}}}\\
\colhead{}  & \colhead{Type}  &\colhead{Composition}  &\colhead{K}
& \colhead{Age, Gyr} & \colhead{pc} & \colhead{km~s$^{-1}$}\\
}
\startdata
ULAS J0121$-$00 & DC & unconstrained & 4480$\pm$230 & 7.3$\,^{+\,2.0}_{-\,3.7}$ & 81$\pm$19 & 38$\pm$10 [51]\\
ULAS J0826$-$00 & DC & H & 4170$\pm$130 & 8.6$\,^{+\,1.3}_{-\,2.6}$ & 76$\pm$10 & 33$\pm$9 [46]\\
ULAS J0840$+$05 & DC & mixed & 4350$\pm$110 & 7.5$\,^{+\,0.4}_{-\,3.0}$ & 82$\pm$15 & 44$\pm$9 [50]\\
ULAS J1006$+$09 & DC & He & 4260$\pm$100 & 7.7$\,^{+\,0.4}_{-\,3.2}$ & 73$\pm$17 & 51$\pm$12 [50]\\
ULAS J1206$+$03 & DC & He & 4610$\pm$80 & 6.9$\,^{+\,0.8}_{-\,3.2}$ & 94$\pm$20 & 73$\pm$15 [71]\\
SDSS J1247$+$06{\tablenotemark{e}} & DQpec & mixed & 5120 & 5.6 & 60 & 110 \\
ULAS J1320$+$08 & DC & high-He mix & 4800$\pm$130 & 6.5$\,^{+\,0.8}_{-\,3.0}$ & 102$\pm$18 & 95$\pm$19 [101]\\
ULAS J1323$+$12 & DA & H & 5260$\pm$90 & 4.3$\,^{+2.9}_{-\,2.1}$ & 103$\pm$21 & 163$\pm$30 [102]{\tablenotemark{f}}\\
ULAS J1345$+$15 & DC & H & 4120$\pm$150 & 8.7$\,^{+1.4}_{-\,2.6}$ & 58$\pm$8 & 72$\pm$10 [71]\\
ULAS J1351$+$12& DC & He & 4810$\pm$70 & 6.5$\,^{+0.9}_{-\,3.2}$ & 103$\pm$21 & 59$\pm$12 [51]\\
ULAS J1404$+$13 & DC & mixed & 4380$\pm$100 & 7.5$\,^{+\,0.4}_{-\,2.9}$ & 59$\pm$10 & 43$\pm$8 [40]\\
ULAS J1436$+$05{\tablenotemark{g}} & DC & unconstrained  & 4340$\pm$170 & 7.8$\,^{+2.0}_{-\,3.7}$ & 57$\pm$12 & 
85$\pm$15 [81]\\
ULAS J1454$-$01 & DC & He & 4800$\pm$60 & 6.5$\,^{+0.9}_{-\,3.2}$& 65$\pm$15 & 65$\pm$15 [83] \\
ULAS J2331$+$15& DA & H & 4970$\pm$80 & 5.8$\,^{+3.7}_{-\,3.1}$ & 95$\pm$27 & 71$\pm$20 [67]\\
\enddata
\tablenotetext{a}{The uncertainty in  $T_{\rm eff}$ is due to both photometric scatter and an allowed range in
gravity of $7.7 \leq \log g \leq 8.3$.}
\tablenotetext{b}{The cooling age is derived from the composition and temperature using the cooling models of Fontaine, Brassard \& Bergeron (2001). The uncertainty is due to the range in gravity of   $7.7 \leq \log g \leq 8.3$. }
\tablenotetext{c}{The distance is estimated from the modelled and observed 
magnitudes (Holberg \& Bergeron 2006). The uncertainty is due to the range in gravity of   $7.7 \leq \log g \leq 8.3$.}
\tablenotetext{d}{The tangential velocity is calculated from distance and the
proper motion given in Table 1. Values in brackets are those implied by the alternative proper motion given in Table 1.}
\tablenotetext{e}{Previously discovered in SDSS by Kilic et al. (2010a), tabulated properties are from that work.}
\tablenotetext{f}{The alternative value for $v_{\rm tan}$ is preferred, given the large uncertainty in the SDSS-LAS
proper motion caused by the short timeline.}
\tablenotetext{g}{The fit to this white dwarf is poor (see Figure 11).}
\end{deluxetable}

\clearpage


\end{document}